\documentstyle[11pt,newpasp,twoside,epsf]{article}
\begin{document}
\title{Computational AstroStatistics: Fast Algorithms and Efficient Statistics
for Density Estimation in Large Astronomical Datasets}

\author{R. C. Nichol,} 
\affil{Department of Physics, Carnegie Mellon University,
5000 Forbes Avenue, Pittsburgh, PA-15213}

\author{A. J. Connolly,} 
\affil{Department of Physics \& Astronomy, University of Pittsburgh} 

\author{A. W. Moore, J. Schneider,} 
\affil{Robotics Institute \& the Computer Science
Department, Carnegie Mellon University, 5000 Forbes Avenue,
Pittsburgh, PA-15213} 

\author{C. Genovese, L. Wasserman,} 
\affil{
Department of Statistics, Carnegie Mellon University, 5000 Forbes Avenue,
Pittsburgh, PA-15213} 

\begin{abstract}
We present initial results on the use of Mixture Models for density
estimation in large astronomical databases. We
provide herein both the theoretical and experimental background for using a
mixture model of Gaussians based on the Expectation Maximization (EM)
Algorithm. Applying these analyses to simulated data sets we show that the EM
algorithm -- using the both the AIC \& BIC penalized likelihood to score the
fit -- can out-perform the best kernel density estimate of the distribution
while requiring no ``fine--tuning'' of the input algorithm parameters. We find
that EM can accurately recover the underlying density distribution from point
processes thus providing an efficient adaptive smoothing method for
astronomical source catalogs. To demonstrate the general application of this
statistic to astrophysical problems we consider two cases of density
estimation; the clustering of galaxies in redshift space and the clustering of
stars in color space. From these data we show that EM provides an adaptive
smoothing of the distribution of galaxies in redshift space (describing
accurately both the small and large-scale features within the data) and a
means of identifying outliers in multi--dimensional color--color space (e.g.\
for the identification of high redshift QSOs). Automated tools such as those
based on the EM algorithm will be needed in the analysis of the next
generation of astronomical catalogs (2MASS, FIRST, PLANCK, SDSS) and
ultimately in in the development of the National Virtual Observatory.
\end{abstract}

\section{Introduction}

With recent technological advances in wide field survey astronomy it has now
become possible to map the distribution of galaxies and stars within the local
and distant Universe across a wide full spectral range (from X-rays through to
the radio) {\it e.g.}  FIRST, MAP, Chandra, HST, ROSAT, SDSS, 2dF, Planck. In
isolation, the scientific returns from these new data sets will be enormous,
combined however, these data will represent a new paradigm for how we
undertake astrophysical research. They present the opportunity to create a
``Virtual Observatory'' that will enable a user to seamlessly analyze and
interact with a multi-frequency digital map of the sky.  While the wealth of
information contained within these new surveys is clear, the questions we must
address is how to we efficiently analyze these massive and intrinsically
multidimensional data sets.  Standard statistical approaches do not easily
scale to the regime of 10$^8$ data points and 100's of dimensions.

Therefore, we have formed a collaboration of computer scientists, statisticans
and astrophysics to address this problem through the development of fast and
efficient statistical algorithms that scale to many dimensions and large
datasets like those found in astronomy. In this volume, we present outlines of
our on--going research including the use of tree algorithms for fast n--point
correlation functions (Schneider et al.), non-parametric density
estimations (Genovese et al.) and data visualisation (Welling et
al.).  In this paper, we present initial results from the application of
Mixture Models to the general problem of density estimation in astrophysics;
the reader is referred to Connolly et al.  (2000, in preparation) for the full
details of the theory, the algorithm and our results.

\section{Outline of the Algorithm}
Let $X^n = (X_1, \ldots , X_n)$ represent the data.  Each $X_i$ is a
$d$-dimensional vector giving, for example, the location of the $i^{th}$
galaxy.  We assume that $X_i$ takes values in a set $A$ where $A$ is a patch
of the sky from which we have observed.  We regard $X_i$ as having been drawn
from a probability distribution with probability density function $f$.  This
means that $f\geq 0$, $\int_A f(x) dx =1$ and the proportion of galaxies in a
region $B$ is given by $Pr (X_i \in B) = \int_B f(x) dx.$
In other words, $f$ is the the normalized galaxy density and the
proportion of galaxies in a region $B$ is just the integral of $f$
over $B$.  Our goal is to estimate $f$ from the observed data $X^n$.

\begin{figure}[tp]
\plottwo{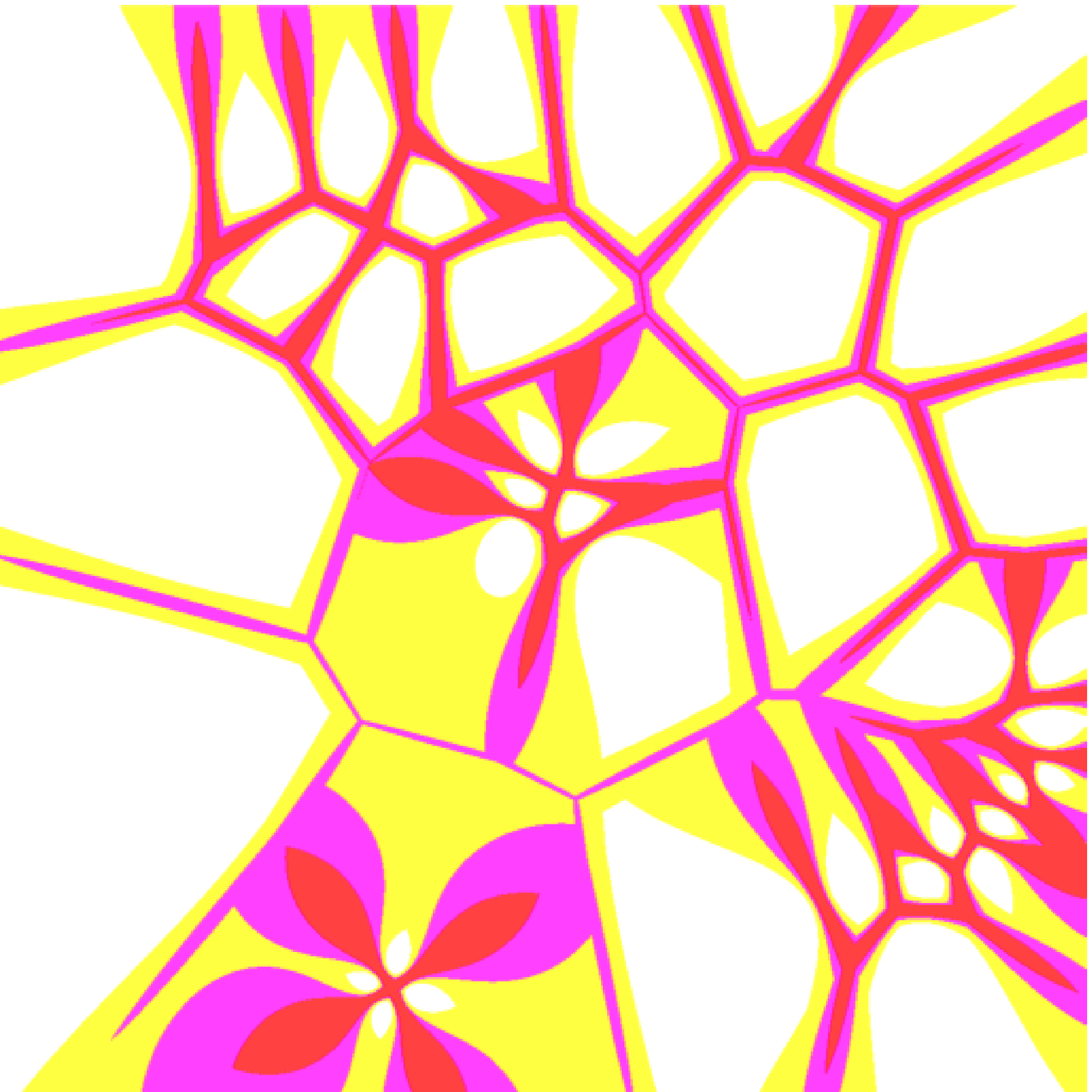}{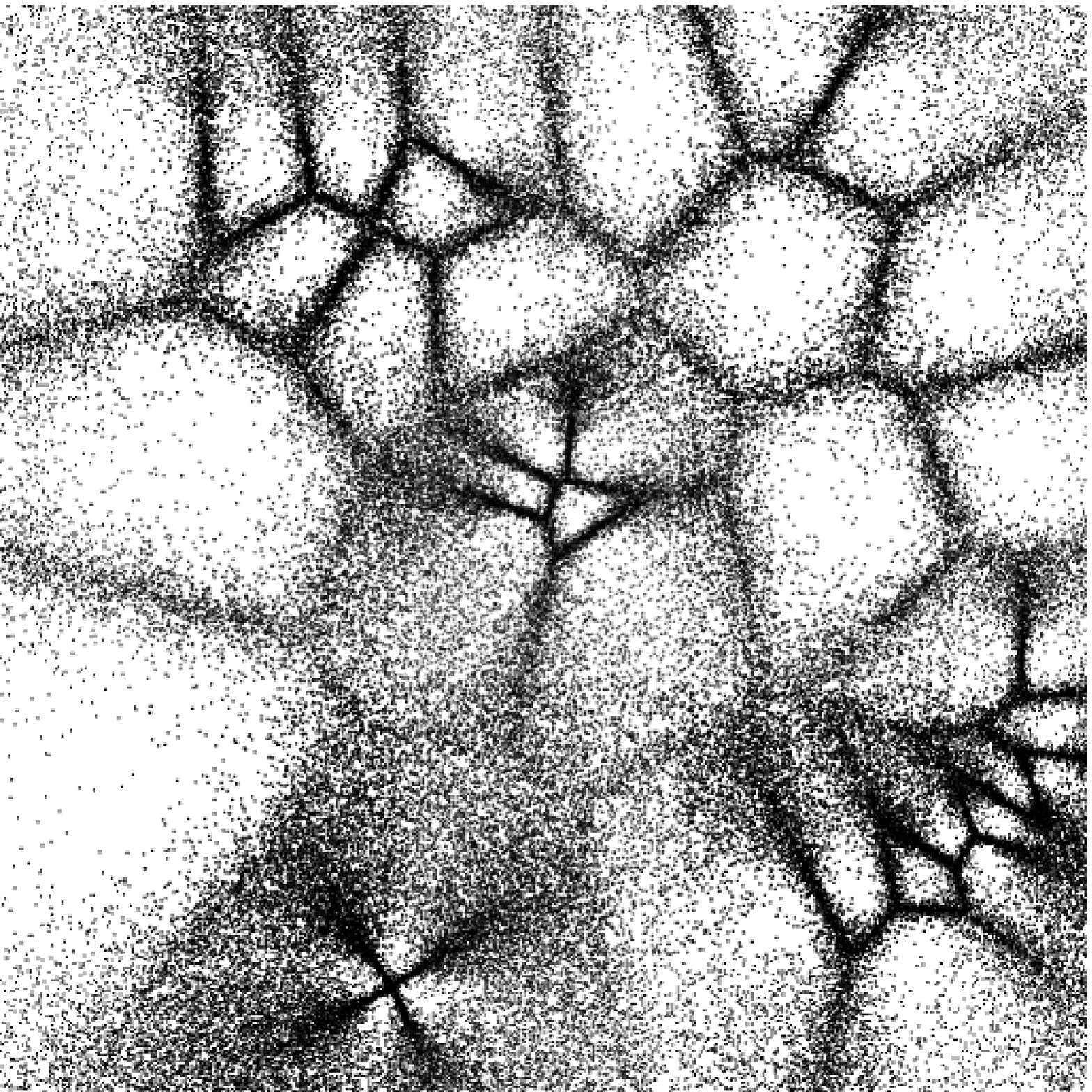}
\caption{\small A comparison of the fixed kernel and our EM density
estimation. Top left panel is the simulated
Voronoi density distribution while top right shows 100,000 data
points draw from this distribution. }
\end{figure}

The density $f$ is assumed to be of the form
\begin{equation}
f(x;\theta_k) = p_0 U(x) +\sum_{j=1}^k p_j \phi(x ; \mu_j, \Sigma_j) 
\end{equation}
where $\phi(x ; \mu, \Sigma)$ denotes a $d$-dimensional Gaussian with
mean $\mu$ and covariance $\Sigma$:
$$
\phi(x ; \mu, \Sigma)=
\frac{1}{ (2\pi)^{d/2} |\Sigma|^{1/2}}
\exp\left\{ -\frac{1}{2} (x-\mu)^T \Sigma^{-1} (x-\mu)\right\},
$$
and $U(\cdot )$ is a uniform density over $A$ i.e.\ $U(x) = 1/V$ for all $x\in
A$, where $V$ is the volume of $A$.  The unknown parameters in this model are
$k$ (the number of Gaussians) and $\theta_k = (p, \mu, \Sigma)$ where $p=(p_0,
\ldots , p_k)$, $\mu = (\mu_1, \ldots , \mu_k)$ and $\Sigma = (\Sigma_1,\ldots
, \Sigma_k)$.  Here, $p_j \geq 0$ for all $j$ and $\sum_{j=0}^k p_j=1$.  This
model is called a mixture of Gaussians (with a uniform component).  The
parameter $k$ controls the complexity of the density $f$.  Larger values of
$k$ allow us to approximate very complex densities $f$ but also entail
estimating many more parameters.  It is important to emphasize that we are not
assuming that the true density $f$ is exactly of the form (Eqn. 1). It
suffices that $f$ can be approximated by such a distribution of Gaussians.
For large enough $k$, nearly any density can be approximated by such a
distribution (see Roeder \& Wasserman 1997).

\begin{figure}[tp]
\plottwo{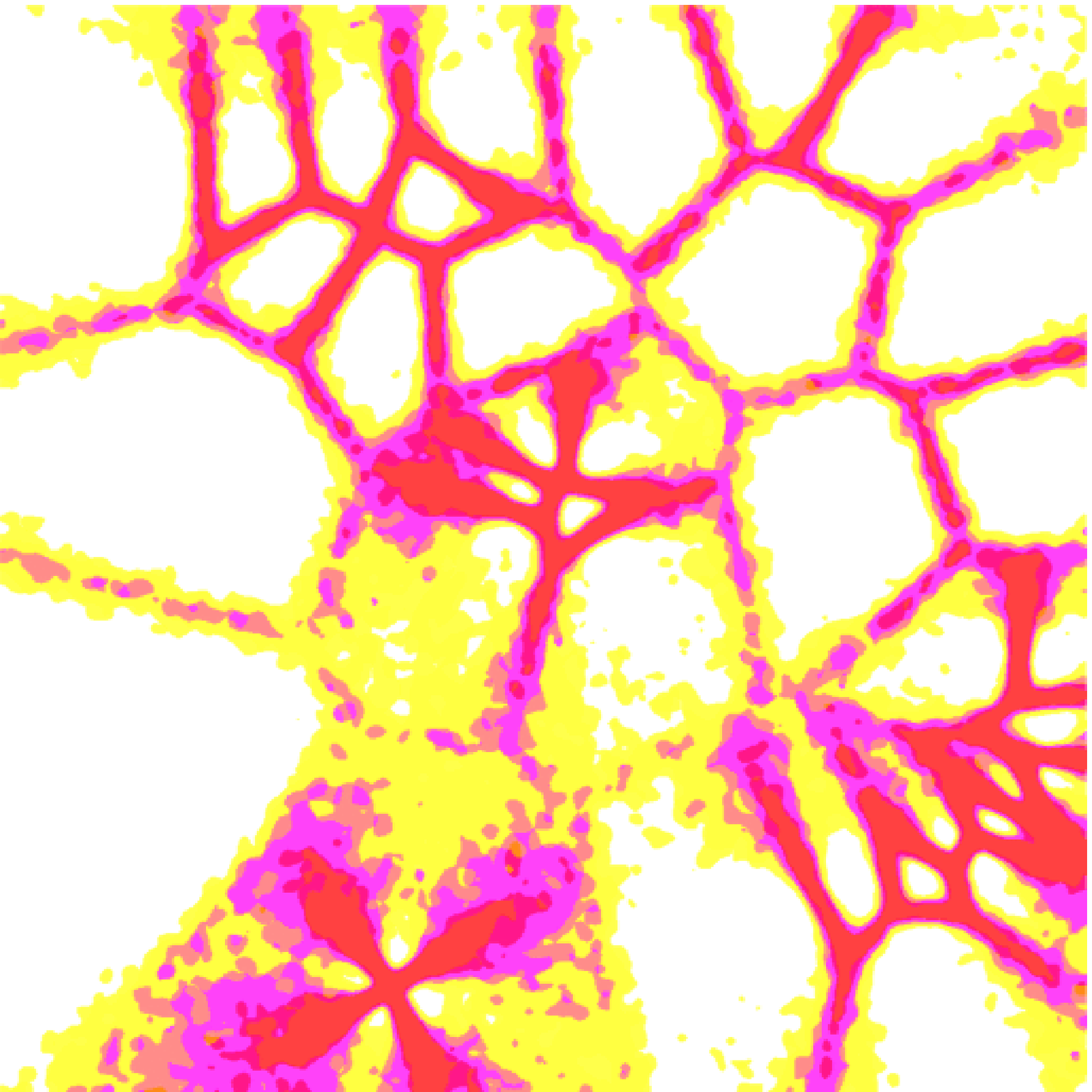}{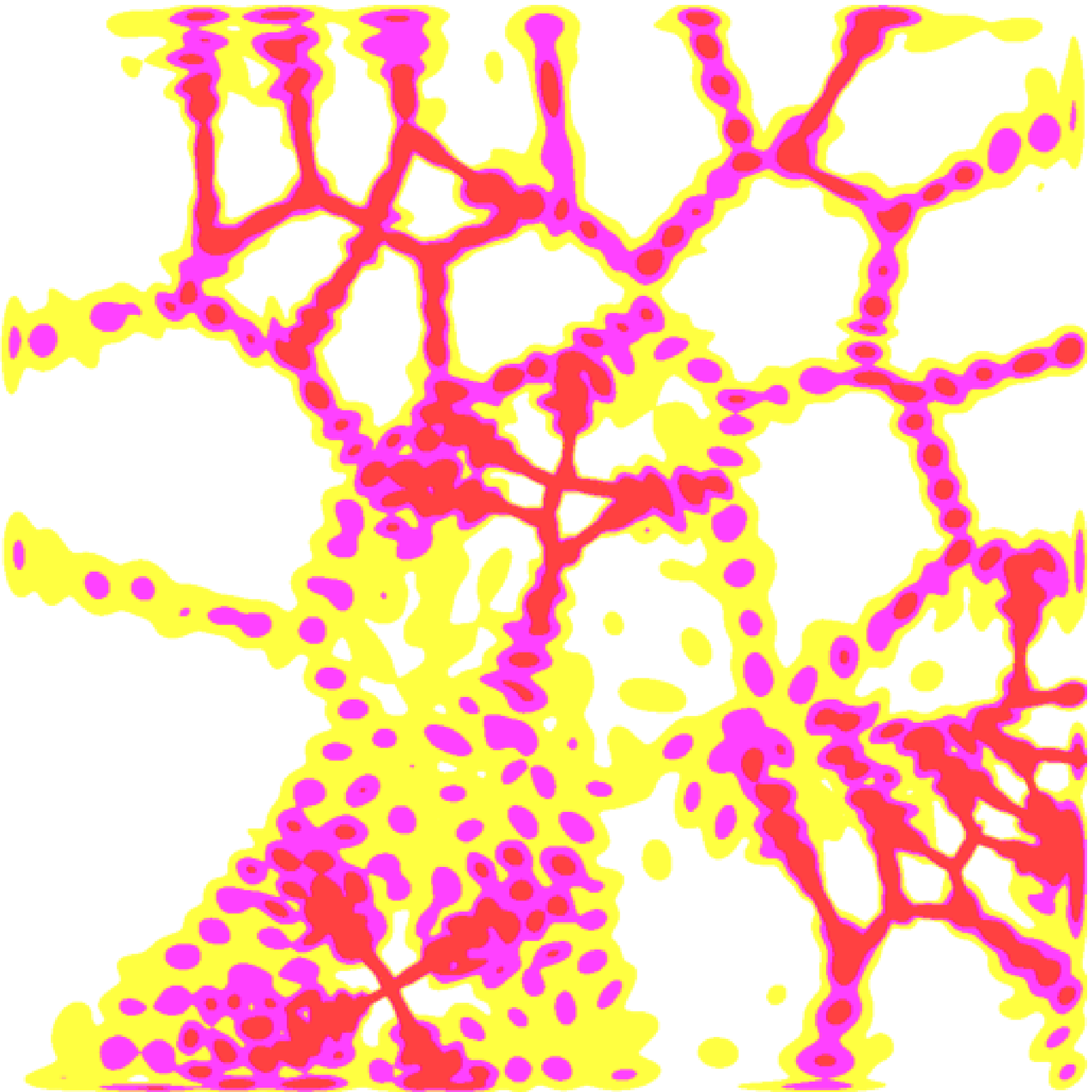}
\caption{\small The resulting density estimates from the data in Figure 1:
fixed kernel (left) and EM+AIC (left)}
\end{figure}

\subsection{The EM Algorithm}

\begin{figure}[t]
\plotone{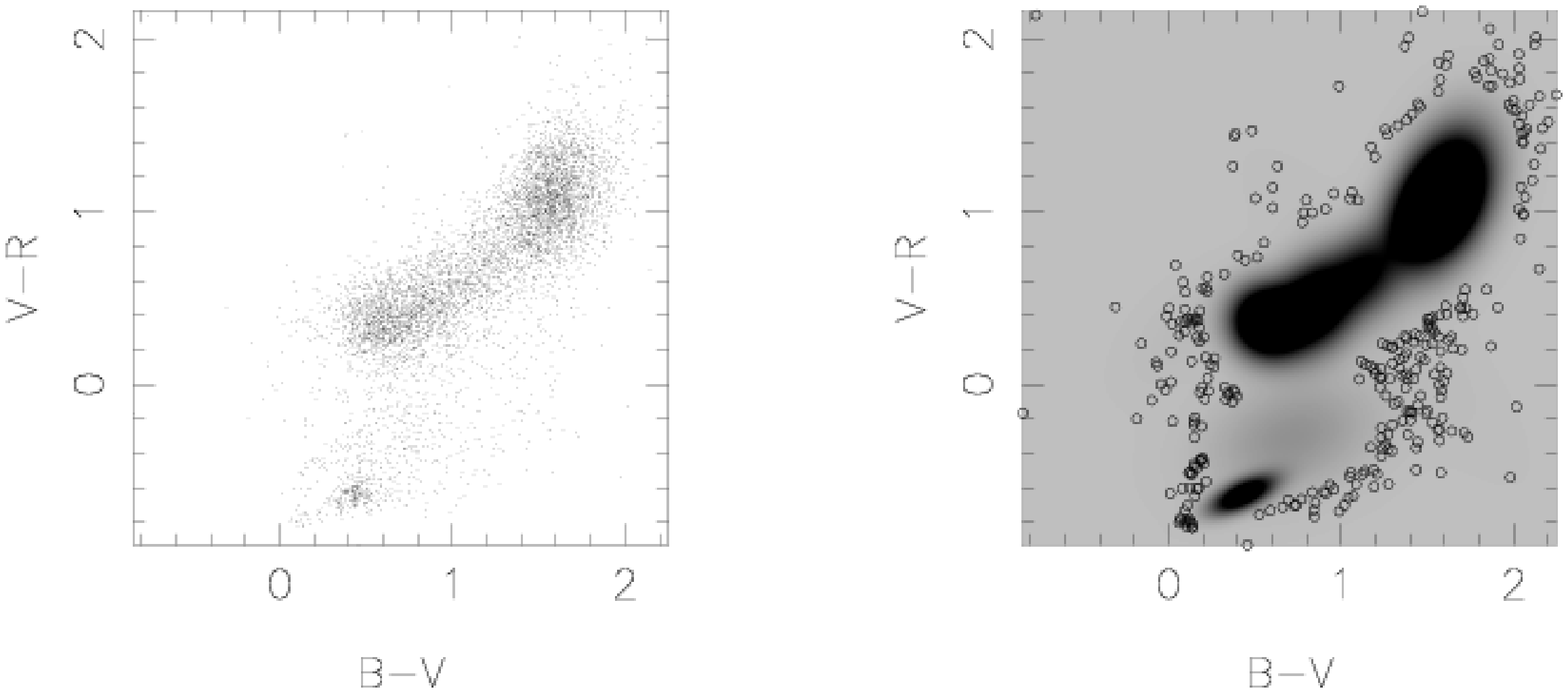}
\caption{\small The application of our algorithm to density estimation in color
space. The left panel shows the distribution of $B-V$ and $V-R$ colors of
stellar sources while the right hand panel shows the density distribution, as
a grayscale, of these sources derived from our algorithm (using AIC).  The
open circles are the 5\% of the data that is least likely to be drawn from
this distribution.}
\end{figure}

How do we find the maximum likelihood estimate of $\theta_k$ {\it i.e.}
$\hat{\theta}$?  (We assume for the moment that $k$ is fixed and known.)  The
usual method for finding $\hat{\theta}$ is the EM (expectation maximization)
algorithm.  We can regard each data point $X_i$ as arising from one of the $k$
components of the mixture.  Let $G=(G_1, \ldots , G_k)$ where $G_i=j$ means
that $X_i$ came from the $j^{th}$ component of the mixture.  We do not observe
$G$ so $G$ is called a latent (or hidden) variable.  Let $\ell_c$ be the
log-likelihood if we knew the latent labels $G$.  The function $\ell_c$ is
called the complete-data log-likelihood.  The EM algorithm proceeds as
follows.  We begin with starting values $\theta_0$. We compute $Q_0 = E(
\ell_c; \theta_0)$, the expectation of $\ell_c$, treating $G$ as random, with
the parameters fixed at $\theta_0$. Next we maximize $Q_0$ over $\theta$ to
obtain a new estimate $\theta_1$. We then compute $Q_1 = E(\ell_c ; \theta_1)$
and continue this process. Thus we obtain a sequence of estimates $\theta_0,
\theta_1, \ldots$ which are known to converge to a local maximum of the
likelihood. See Connolly et al. (2000) for more details.

Above, we have assumed that $k$ is known.  One approach of choosing $k$ from
the data is to sequentially test a series of hypotheses i.e. test the density
estimation using $k$ versus the one with $k+1$ and repeat for various $k$.The
usual test for comparing such hypotheses is called the ``likelihood ratio
test'' which compares the value of the maximized log-likelihood under the two
hypotheses.  This approach is infeasible for large data sets where $k$ might
be huge.  Also, this requires knowing the distribution of the likelihood ratio
statistic which is not known. As an alternative (see Connolly et al. 2000 for
the explanation), we use two common penalized log--likelihoods to test these
two hypotheses, namely Akaike Information Criterion (AIC) and Bayesian
Information Criterion (BIC) which take the general form of
$\ell(\hat{\theta}_k) - \lambda_k R_k$ where $R_k$ is the number of parameters
in the $k^{th}$ model. For AIC $\lambda_k=1$ while $\lambda_k=\log n /2$ gives
the BIC criterion. See Connolly et al. (2000) for a full explanation.

Given the definition of the EM algorithm and the criteria for determining the
number of components to the model using AIC and BIC we must now address how do
we apply this formalism to massive multi--dimensional astronomical
datasets. In its conventional implementation, each iteration of EM visits
every data point pair, meaning $kR$ evaluations of a $M$-dimensional Gaussian,
where $R$ is the number of data-points and $k$ is the number of components in
the mixture. It thus needs $O(M^2 kR)$ arithmetic operations per
iteration. For data sets with 100s of attributes and millions of records such
a scaling will clearly limit the application of the EM technique. We must
therefore develop an algorithmic approach that reduces the cost and number of
operations required to construct the mixture model. To do this, we have used
Multi-resolution KD-trees to gain impressive speed--ups for the numerous
range--searches involved in computing the fits of these $k$ gaussians to the
data. Such tree algorithms are discussed in detail by Schneider et al.
in this volume.

\section{Applications to Astrophysical Problems}

To test the sensitivity of the Mixture Model density estimation to the
heirarchical clustering in the universe, and thus determine if it is better
than present, more traditional, astronomical methods of density estimations,
we have tested our algorithm using simulated data generated from a Voronoi
Tessellation since this mimics the observed distribution of filaments and
sheets of galaxies in the universe.  In Figure 1, we present the underlying
Voronoi density map we have constructed; this is the ``truth'' in our
simulation. From this distribution we derive a set of 100,000 data points to
represent a mock 2-dimensional galaxy catalog.  

We have applied the EM algorithm (with both the AIC and BIC criteria) and a
standard fixed kernel density estimator to these point-like data sets in order
to reproduce the original density field.  The latter involved finely binning
the data and smoothing the subsequent grid with a binned Gaussian filter of
fixed bandwidth which was chosen {\bf by hand} to minimize the
Kullback-Leibler distance ($D(f,g) = \int f(x) \log f(x) /g(x)dx$) between the
resulting smoothed map and the true underlying density distribution. Clearly,
we have taken the optimal situation for the fixed kernel estimator since we
have selected it's bandwidth to ensure as close a representation of the true
density distribution as possible. This would not be the case in reality.

In Figure 2, we show the reconstructed density field using the fixed kernel
and EM algorithm (AIC).  As we would expect the fixed kernel technique
provides an accurate representation of the overall density field. The kernel
density map suffers, however, when we consider features that are thinner than
the width of the kernel. Such filamentary structures are over-smoothed and
have their significance reduced. In contrast the EM algorithm attempts to
adapt to the size of the structures it is trying to reconstruct. The right
panel shows that where narrow filamentary structures exist the algorithm
places a large number of small Gaussians. For extended, low frequency,
components larger Gaussians are applied.  For the fixed kernel estimator, we
have a measured KL divergence of 0.074 between the final smoothed map and the
true underlying density map (remember, this is the smallest KL measurement by
design).  For the EM AIC density map we measure a KL divergence of 0.067 which
is lower than the best fixed kernel KL score thus immediately illustrating the
power of the EM methodology. We have not afforded the same prior knowledge to
the EM measurement -- {\it i.e.\ } hand--tune it so as to minimize the KL
divergence -- yet we have beaten the kernel estimator.

We have extended the analysis above to real astronomical data {\it e.g.} the
the distribution of 6298 stellar sources in the B-V and V-R color space (with
R$<$22) taken from a 1 sq degree multicolor photometric survey of Szokoly et
al. (2000).  In this case, the mixture model density distribution naturally
provides a probability density map that a stellar object drawn at random from
the observed distribution of stars would have a particular set of B-V -- V-R
colors. We can now assign a probability to each star in the original data that
describes the likelihood that it arises from the overall distribution of the
stellar locus. We can then rank order all sources based on the likelihood that
they were drawn from the parent distribution. The right panel of Figure 3
shows the colors of the 5\% of sources with the lowest probabilities. These
sources lie preferentially away from the stellar locus and thus can be used to
find high redshift quasars as outliers to color--space. As we increase the cut
in probability the colors of the selected sources move progressively closer to
the stellar locus.

The advantage of the EM approach over standard color selection techniques is
that we identify objects based on the probability that they lie away from the
stellar locus (i.e.\ we do not need to make orthogonal cuts in color space as
the probability contours will trace accurately the true distribution of
colors). While for two dimensions this is a relatively trivial statement as it
is straightforward to identify regions in color--color space that lie away
from the stellar locus (without being restricted to orthogonal cuts in
color-color space) this is not the case when we move to higher dimensional
data.  For four and more colors we lose the ability to visualize the data with
out projecting it down on to a lower dimensionality subspace (i.e.\ we can
only display easily 3 dimensional data). In practice we are, therefore,
limited to defining cuts in these subspaces which may not map to the true
multidimensional nature of the data. The EM algorithm does not suffer from
these disadvantages as a probability density distribution can be defined in an
arbitrary number of dimensions. It, therefore, provides a more natural
description of both the general distribution of the data and for the
identification of outlier points from high dimensionality data sets. With the
new generation of multi-frequency surveys we expect that the need for
algorithms that scale to a large number of dimensions will become more
apparent.

\end{document}